\documentclass[twocolumn,preprintnumbers,showpacs,showkeys,amsmath,amssymb,epsf]{revtex4}
\usepackage{graphicx}
\usepackage{dcolumn}
\usepackage{bm}
\usepackage{amsfonts}
\usepackage{amssymb}
\usepackage{bbm}
\usepackage{bbold}
\usepackage{color}
\begin{document}
\title{Enantiodromic effective generators of a Markov jump process with Gallavotti-Cohen symmetry}
\author{S. A. A. Terohid}
\email{a.terohid@iauh.ac.ir}
\author{P. Torkaman}
\email{p.torkaman@basu.ac.ir}
\author{F. H. Jafarpour}
\email{farhad@ipm.ac.ir}
\affiliation{Physics Department, Bu-Ali Sina University, 65174-4161 Hamedan, Iran}
\date{\today}
\begin{abstract}
This paper deals with the properties of the stochastic generators of the effective (driven) processes associated with atypical values of 
transition-dependent time-integrated currents with Gallavotti-Cohen symmetry in Markov jump processes. Exploiting the concept of biased 
ensemble of trajectories by introducing a biasing field $s$, we show that the stochastic generators of the effective processes associated with 
the biasing fields $s$ and $E-s$ are enantiodromic with respect to each other where $E$ is the conjugated field to the current. We 
illustrate our findings by considering an exactly solvable creation-annihilation process of classical particles with 
nearest-neighbor interactions defined on a one-dimensional lattice. 
\end{abstract}
%%%%%%%%%%%%%%%%%%%%%%%%%%%%%%%%%%%%%%%%%%%%%%%%%%%%%%%%%%%%%%%%%%%%%%
\maketitle
%%%%%%%%%%%%%%%%%%%%%%%%%%%%%%%%%%%%%%%%%%%%%%%%%%%%%%%%%%%%%%%%%%%%%%

%%%%%%%%%%%%%%%%%%%%%%%%%%%%%%%%%%%%%%%%%%%%%%%%%%%%%%%%%%%%%%%%%%%%%%
\section{Introduction}
%%%%%%%%%%%%%%%%%%%%%%%%%%%%%%%%%%%%%%%%%%%%%%%%%%%%%%%%%%%%%%%%%%%%%%
The appearance of rare events in equilibrium and non-equilibrium many-body systems have become 
a focus of recent intense research~\cite{AF01,PF,AJK06}. An important question is how these fluctuations arise. 
In order to answer this question let us consider a classical interacting particles system 
in its steady-state. We assume that this system can be modeled by a Markov jump process in continuous time in or 
out of equilibrium~\cite{S01}. In a Markov jump process the system jumps spontaneously from one classical 
configuration in the configuration space, which will be assumed to be finite-dimensional throughout this paper, to another classical configuration 
with certain transition rate making a trajectory or path. We are generally interested in measuring a transition-dependent time-integrated observable 
such as activity or particle current during an extended period of time called the observation time. As we mentioned, we are more specifically 
interested in the fluctuations of this observable.  
Imagine that the stationary probability distribution function of this observable can be built by doing some experiments and measuring the 
observable along the trajectories of the process in its steady-state. In principle, both typical and atypical values (or fluctuations) of the 
observable can be observed. Some of these trajectories are responsible for creating the typical values of the observable while some other 
trajectories are responsible for creating a specific fluctuation or an atypical value of the observable. Now, if we look 
at a restricted set of trajectories which is responsible for a specific fluctuation, we are basically dealing with a conditioning. Here the corresponding 
ensemble is called the path microcanonical ensemble. There are actually many trajectories leading to a specific fluctuation and one might ask if it 
is possible to describe this by a stochastic Markov process. It has been shown that in the long observation-time limit each specific fluctuation can be described 
by a specific stochastic Markov process which is non-conditioned and is called the effective or driven process~\cite{JS10}. 

The stochastic generator associated with the effective process for a specific fluctuation can, in principle, be obtained as follows. We start with 
a so called tilted generator~\cite{LAV07,S09,PSS10,BS13,BS2013,HPS13,HMS15,S15}. The tilted generator can be constructed by 
multiplying the transition rates of the original stochastic Markov process by an exponential factor which depends on both a biasing 
field $s$ and the increment of the current during that transition. The parameter $s$ is a real parameter 
which selects the fluctuation and plays a role similar to the inverse of temperature in the ordinary statistical mechanics.
Given that our observable satisfies the large deviation principle with a convex rate function and that the tilted generator 
has a spectral gap, it has been shown that using a generalization of Doob's h-transform of the tilted generator one can 
obtain the stochastic generator of the effective process which explicitly depends on $s$. The path ensemble associated with 
the effective dynamics is called the path canonical ensemble. It has already been shown that the microcanonical and the canonical
path ensembles are equivalent in the long-time limit~\cite{CT2}. It is worth mentioning that the effective process 
inherits many properties of the original process including the symmetries; however, the interactions might be very complicated, hence 
the characterization of them are of great importance~\cite{JS15}. A one-dimensional classical Ising chain which exhibits ferromagnetic 
ordering in its biased ensemble of trajectories, is an example which reveals this feature~\cite{JS10}. Similar examples are also 
studied in~\cite{JS14,PS11}. 

The question we are aiming to answer in this paper is that, for a given transition-dependent time-integrated current, whether the 
stochastic generators of the effective processes associated with two different fluctuations are related. Let us start with 
a non-equilibrium Markov jump process in continuous time in the steady-state. We also consider a transition-dependent time-integrated 
current as an observable which satisfies the large deviation principle with a convex rate function~\cite{T09}. Now we assume that 
this non-equilibrium system is created as a results of applying an external field $E$ conjugated to that current to an equilibrium process. 
Among different choices, one can achieve this by a special scaling of the reaction rates of the equilibrium system. 
This will clearly restrict us to study only a very specific class of processes. Moreover, this specific type of scaling results in a 
rate function which satisfies the Gallavotti-Cohen symmetry with respect to $E$~\cite{GCEM}. 
We will show that in a given Markov jump process the stochastic generator of the effective process at $s$ is {\em enantiodromic} with 
respect to the stochastic generator of the effective process at $E-s$. The concept of the enantiodromy relation has already been introduced 
and used in the field of stochastic interacting particles~\cite{S01,SH04}. In order to show how this property helps us deduce 
more information about the effective interactions associated with certain fluctuations in the system, we will provide an exactly solvable example. 
Our example consists of a one-dimensional system of classical particles with nearest-neighbour interactions which includes 
creation and annihilation of particles. As we will see, using the enantiodromy relation and knowing the effective 
interactions at $s$, one finds exact information about the effective interactions at $E-s$. It turns out that the 
nature of interactions can be quite different from each other at these two points.   

This paper is organized as follows: in the second section, after a brief review of the basic mathematical concepts, we will 
bring the main results. The third section is devoted to an application of our findings by providing an exactly solvable 
coagulation-decoagulation model of classical particles. In the last section we will bring the concluding remarks. 

%%%%%%%%%%%%%%%%%%%%%%%%%%%%%%%%%%%%%%%%%%%%%%%%%%%%%%%%%%%%%%%%%%%%%%
\section{Basics concepts and results}
%%%%%%%%%%%%%%%%%%%%%%%%%%%%%%%%%%%%%%%%%%%%%%%%%%%%%%%%%%%%%%%%%%%%%%
Let us start with a Markov process in continuous time defined by a set of configurations $ \{c\}$ and transition 
rates $\omega_{c \to c'}$ between its configurations in a finite-dimensional configuration space. Considering the 
vectors $\{ | c \rangle \}$ as an orthogonal basis of a complex vector space, the probability of being in configuration $c$ at 
time $t$ is given by $P(c,t)=\langle c | P(t) \rangle$. The time evolution of $| P(t) \rangle$ is governed  by the following 
master equation~\cite{S01}
\begin{equation}
\label{ME}
\frac{d}{dt} | P(t) \rangle = \hat{\cal{H}} | P(t)\rangle
\end{equation}
in which the stochastic generator or Hamiltonian $\hat{\cal{H}}$ is a square matrix with the following matrix elements
\begin{equation}
\label{HME}
\langle c' | \hat{\cal{H}} | c \rangle =(1-\delta_{c,c'}) \omega_{c \to c'} - \delta_{c,c'} \sum_{c'' \neq c} \omega_{c \to c''} \; .
\end{equation}
We assume that in the long-time limit the process reaches its steady-state so that the left-hand side of~(\ref{ME}) becomes 
zero. We aim to study the fluctuations of an observable over a long observation time in the steady-state. 
As a dynamical observable we consider a transition-dependent time-integrated current. The current is time-extensive and  
a functional of the trajectory that the system follows in the configuration space during the observation time. This is a sum 
of the increments $\theta_{c\to c'}$'s every time a jump from $c$ to $c'$ occurs. For particle current in one-dimension we 
have $\theta_{c\to c'}=\pm 1$. 

Let us now assume that our original process has a non-equilibrium steady-state which has been obtained by applying 
an external driving field to an equilibrium process. Among different possibilities, a very special choice for making this connection 
is by considering the following rule
\begin{equation}
\label{MCON}
\omega_{c\to c'} =\omega^{\text {eq}}_{c \to c'}e^{\frac{E}{2}\theta_{c \to c'}} 
\end{equation}
in which $E$ is the external driving field conjugated to the current. The transition rates of the equilibrium process and the 
equilibrium stationary distribution satisfy the detailed balance equations i.e.
$\omega^{\text {eq}}_{c \to c'} P_{\text eq}(c)=\omega^{\text {eq}}_{c' \to c} P_{\text eq}(c')$
which can be written as
\begin{equation}
\label{LDB}
\omega_{c\to c'} e^{-\frac{E}{2} \theta_{c \to c'}}P_{\text{eq}}(c)=\omega_{c'\to c} e^{-\frac{E}{2} \theta_{c' \to c}}P_{\text{eq}}(c') \; .
\end{equation}
Defining the modified Hamiltonian $\hat{\cal{H}}(s)$ (or the tilted generator) with the 
matrix elements~\cite{PSS10,BS13,BS2013,HPS13}
\begin{equation}
\label{MHME}
\langle c' | \hat{\cal{H}}(s) | c \rangle = (1-\delta_{c,c'})e^{-s {\theta}_{c \to c'}} \omega_{c \to c'} - \delta_{c,c'} 
\sum_{c'' \neq c} \omega_{c \to c''} 
\end{equation}
one can write~(\ref{LDB}) in a matrix form
\begin{equation}
\label{LDB2}
\hat{\cal{H}}(E-s)=P_{\text{eq}}\hat{\cal{H}}^{\text{T}}(s)P^{-1}_{\text{eq}}
\end{equation}
in which $P_{\text{eq}}$ is a diagonal matrix with the matrix elements 
$\langle c \vert P_{\text{eq}}\vert c \rangle =P_{\text{eq}}(c)$~\cite{HS07}. 
Considering the following eigenvalue equations for the modified Hamiltonian 
\begin{equation}
\label{EVP}
\begin{array}{l}
\hat{\cal{H}}(s) \vert \Lambda (s) \rangle = \Lambda (s) \vert \Lambda  (s) \rangle \; , \\ \\
\langle \tilde{\Lambda}  (s) \vert \hat{\cal{H}}(s)  = \Lambda (s) \langle \tilde{\Lambda}  (s) \vert  
\end{array}
\end{equation}
it is clear from~(\ref{LDB2}) that all of the eigenvalues of the modified Hamiltonian have the following symmetry 
$\Lambda(s)=\Lambda(E-s)$ which includes its largest eigenvalue i.e. 
\begin{equation}
\label{GCSR2}
\Lambda^{\ast}(s)=\Lambda^{\ast}(E-s) 
\end{equation}
which is called the Gallavotti-Cohen symmetry~\cite{GCEM,HS07,LS99}. Finally, the similarity transformation~(\ref{LDB2})   
indicates that
\begin{equation} 
\vert \Lambda (s) \rangle=P_{\text{eq}}\vert \tilde{\Lambda} (E-s) \rangle\; .
\end{equation}

As we have already explained, in the long observation-time limit each specific fluctuation in the system can be described by a stochastic 
Markov process called the effective process which is equivalent to the conditioning of the original process on
seeing a certain fluctuation. It has been shown that the stochastic generator of this effective stochastic process is given by
\begin{equation}
\label{effective relation}
\hat{\cal{H}}_{\text{eff}}(s) =U(s) \hat{\cal{H}}(s) U^{-1}(s)-\Lambda^{\ast}(s) 
\end{equation}
which is a generalization of Doob's h-transform and that $U(s)$ is a diagonal matrix with the matrix element 
$ \langle c | U(s) | c \rangle = \langle \tilde{\Lambda}^{{\ast}}(s) |c \rangle $~\cite{JS10,CT2} . 
The off-diagonal matrix elements of the operator $\hat{\cal{H}}_{\text{eff}}(s)$ in~(\ref{effective relation}) are given by 
\begin{equation}
\label{element}
\langle c' | \hat{\cal{H}}_{\text{eff}}(s) | c \rangle =\langle c' | \hat{\cal{H}}(s) | c \rangle \frac{
 \langle \tilde{\Lambda}^{{\ast}}(s) | c' \rangle}
{ \langle \tilde{\Lambda}^{{\ast}}(s) |c \rangle} 
\end{equation}
or equivalently
\begin{equation}
\label{TR old}
\omega^{\text{eff}}_{c\to c'}(s)=e^{-s\theta_{c\to c'}}\omega_{c\to c'} \frac{
 \langle \tilde{\Lambda}^{{\ast}}(s) | c' \rangle}
{ \langle \tilde{\Lambda}^{{\ast}}(s) |c \rangle}  \; .
\end{equation}
Defining the diagonal matrix 
\begin{equation}
\label{DMD}
U_{\text {TTI}}(s)=\vert  \Lambda^{{\ast}}(s) \rangle \langle \tilde{\Lambda}^{{\ast}}(s)  \vert
\end{equation}
the steady-state distribution of $\hat{\cal{H}}_{\text{eff}}(s)$ is given by  
\begin{equation}
\label{SSDS}
P_{\text{TTI}}(c,s)=\langle c \vert P_{\text{TTI}}(s) \rangle  = \langle c \vert U_{\text {TTI}}(s)\vert c \rangle 
\end{equation}
where the subscript TTI is an abbreviation for the Time Translational Invariance regime~\cite{JS10}. 

Starting from~(\ref{GCSR2}) and using~(\ref{effective relation}) and after some algebra one finds the following enantiodromy relation 
\begin{equation}
\label{enantiodromy 1}
\hat{\cal{H}}_{\text{eff}}(E-s) =  U_{\text {TTI}}(s)       \hat{\cal{H}}_{\text{eff}}^{\text T}(s)       U_{\text {TTI}}^{-1}(s) 
\end{equation}
in which $\text T$ means the transpose of the square matrix. The notion of enantiodromy has its origin in the time-reversal of the 
temporal order, as the transpose matrix describes the motion of the process {\em backward} in time. The two effective stochastic generators 
$\hat{\cal{H}}_{\text{eff}}(s)$ and $\hat{\cal{H}}_{\text{eff}}(E-s)$ share the same spectrum and also the same stationary 
distribution~\cite{S01,HS07}; however, they describe two different processes. For instance while $\hat{\cal{H}}_{\text{eff}}(s)$ 
might contain local and short-range interactions, $\hat{\cal{H}}_{\text{eff}}(E-s)$ can contain long-range and complicated interactions. 
The generator $\hat{\cal{H}}_{\text{eff}}(E-s)$ is called the adjoint generator with respect to $ \hat{\cal{H}}_{\text{eff}}(s)$ which in the 
absence of detailed balance defines a new process with the same allowed transitions as $ \hat{\cal{H}}_{\text{eff}}(s)$~\cite{HS07}.

Using~(\ref{enantiodromy 1}) we find that the transition rates of these effective processes are related through
\begin{equation}
\label{TR new}
\omega^{\text{eff}}_{c\to c'}(E-s)=\omega^{\text{eff}}_{c'\to c}(s)\frac{P_{\text{TTI}}(c',s)}{P_{\text{TTI}}(c,s)} \; .
\end{equation}
It can be seen that the transition rates of the effective process at $E-s$ depend on both the reversed transition rates and the 
stationary distribution of the effective process at $s$. It is worth to mention that~(\ref{TR new}) is obtained by assuming that 
the original process has the Gallavotti-Cohen symmetry in the sense of~(\ref{MCON}). Finally~(\ref{TR new}) gives the following 
constraints on the transition rates of the above mentioned effective processes
\begin{equation}
\omega^{\text{eff}}_{c\to c'}(E-s)\omega^{\text{eff}}_{c'\to c}(E-s)=\omega^{\text{eff}}_{c\to c'}(s)\omega^{\text{eff}}_{c'\to c}(s) \; .
\end{equation}
Similar result has already been obtained in~\cite{E10} for fluids under continuous shear.

Let us consider $s=E/2$. Because of the Gallavotti-Cohen symmetry, this point is the minimum of the largest eigenvalue
$\Lambda^\ast(s)$. At this point the slop of the eigenvalue is zero, hence the average current is zero. This means that the effective process
is in equilibrium. From~(\ref{enantiodromy 1}) one finds
$$
\hat{\cal{H}}_{\text{eff}}(\frac{E}{2}) =  U_{\text {TTI}}(\frac{E}{2})\hat{\cal{H}}_{\text{eff}}^{\text T}(\frac{E}{2})U_{\text {TTI}}^{-1}(\frac{E}{2}) 
$$
which is a self-enantiodromy relation for the stochastic generator of the effective process at $s=E/2$~\cite{S01}. Since at $s=E/2$ the effective 
process is in equilibrium, the detailed-balance condition has to be recovered i.e.
$$
P_{\text{TTI}}(c,\frac{E}{2}) \omega^{\text{eff}}_{c\to c'}(\frac{E}{2})=P_{\text {TTI}}(c',\frac{E}{2}) \omega^{\text{eff}}_{c'\to c}(\frac{E}{2}) \; .
$$
Using~(\ref{TR old}), ~(\ref{DMD}) and ~(\ref{SSDS}) one can readily find
$$
\frac{\omega_{c\to c'} e^{-\frac{E}{2} \theta_{c \to c'}}}{\omega_{c'\to c} e^{-\frac{E}{2} \theta_{c' \to c}}}=
\frac{ \langle \tilde{\Lambda}^{{\ast}}(\frac{E}{2}) | c \rangle \langle c' | \Lambda^{{\ast}}(\frac{E}{2}) \rangle}
{ \langle \tilde{\Lambda}^{{\ast}}(\frac{E}{2}) | c' \rangle \langle c | \Lambda^{{\ast}}(\frac{E}{2}) \rangle} \; .
$$
Comparing this relation with~(\ref{LDB}) we also find
$$
P_{\text{eq}}(c)=\frac{\langle c | \Lambda^{{\ast}}(\frac{E}{2}) \rangle}{\langle \tilde{\Lambda}^{{\ast}}(\frac{E}{2}) | c \rangle} \; .
$$
Note that we have~\cite{JS10} 
$$
P_{\text{TTI}}(c,\frac{E}{2})=\langle c | \Lambda^{{\ast}}(\frac{E}{2}) \rangle \langle \tilde{\Lambda}^{{\ast}}(\frac{E}{2}) | c \rangle \; .
$$
Finally, at $s=0$ the relation~(\ref{TR new}) becomes
\begin{equation}
\label{s=0}
\omega^{\text{eff}}_{c\to c'}(E)=\omega_{c'\to c}\frac{P^{\ast}(c')}{P^{\ast}(c)} \; .
\end{equation}
in which $\omega_{c'\to c}$'s and $P^{\ast}(c)$ are the transition rates and the steady-state probability distribution 
of our original non-equilibrium process.

%%%%%%%%%%%%%%%%%%%%%%%%%%%%%%%%%%The Model%%%%%%%%%%%%%%%%%%%%%%%%%%%%%%
\section{An exactly solvable example}
%%%%%%%%%%%%%%%%%%%%%%%%%%%%%%%%%%The Model%%%%%%%%%%%%%%%%%%%%%%%%%%%%%%
%%%%%%%%%%%%%%%%%%%%%%%
\begin{figure*}
\includegraphics[scale=0.65]{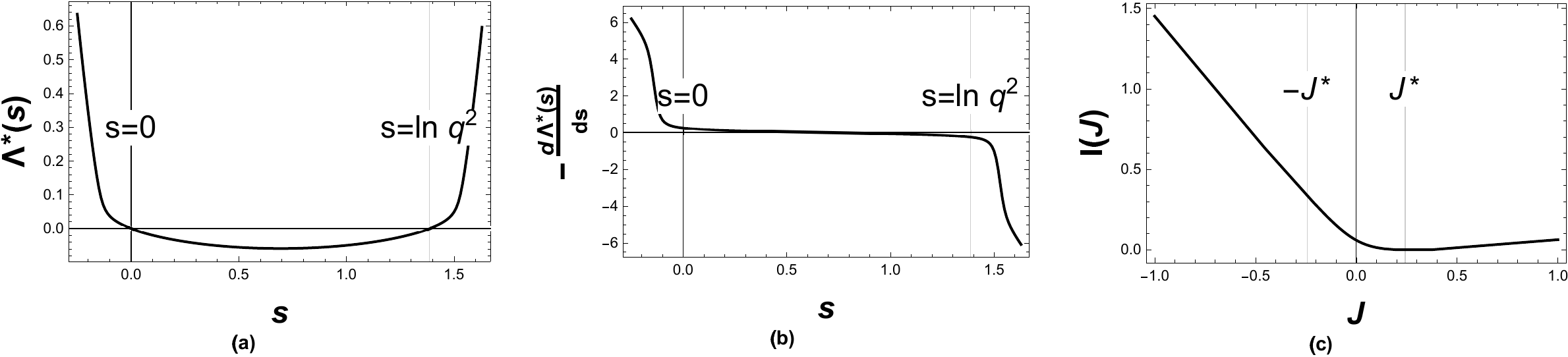}
\caption{\label{fig1} In (a) we have plotted the largest eigenvalue of~(\ref{hs}) and in (b) its first derivative for $L=8$, $q=2$ and $\Delta=0.5$.
For these values of the reaction rates the system is in the low-density phase. We have also plotted the large deviation function of the current $I(J)$ in (c).}
\end{figure*}
%%%%%%%%%%%%%%%%%%%%%%
In what follows we use~(\ref{s=0}) to investigate the effective dynamics at $s=E$ of an exactly solvable coagulation-decoagulation model
of classical particles on a one-dimensional lattice of length $L$ with reflecting boundaries. The system evolves in time according 
to the following reaction rules
\begin{equation}
\label{Rules}
\begin{array}{ll}
A+A \rightarrow \emptyset+A & \mbox{with rate} \; \; q^{-1} \; ,\\
A+\emptyset \rightarrow \emptyset+A & \mbox{with rate} \; \; q^{-1}  \; , \\
A+A \rightarrow A+\emptyset & \mbox{with rate} \; \; q \; , \\
\emptyset+A \rightarrow A+\emptyset & \mbox{with rate} \; \; q \; , \\
\emptyset+A \rightarrow A+A & \mbox{with rate} \; \; \Delta q \; , \\
A+\emptyset \rightarrow A+A & \mbox{with rate} \; \; \Delta q^{-1}  
\end{array}
\end{equation}
where $A$ and $\emptyset$ stand for the presence of a particle and a vacancy at a given lattice-site respectively. There is no input
or output of particles at the boundaries. We assume that $q \ge 1$ and $\Delta > 0$. The steady-state properties of the model have 
been studied in detail using different techniques~\cite{HKP96,HSP96,J04,JM08}. The full spectrum of the stochastic generator and 
also the density profile of the particle in the steady-state have been calculated exactly~\cite{HKP96,HSP96}. The stochastic generator 
of the system is reducible. The system does not evolve in time if it is completely empty. However, there is a non-trivial steady-state 
where there is at least one particle on the lattice.
It is known that in the non-trivial case the system undergoes a phase transition from a high-density phase ($q^2 < 1+\Delta$) into a 
low-density phase ($q^2 > 1+\Delta$) at $q^2=1+\Delta$. Last but not least, it has been shown that the steady-state of this system 
can be written as a linear superposition of shocks which perform random walk on the lattice~\cite{JM08}.  

In~\cite{TJ13} the authors have shown that one can define an entropic reaction-diffusion current in this system with the Gallavotti-Cohen symmetry 
respecting to the conjugate field $E=\ln q^2$. Assigning an occupation number $c_i$ to the lattice-site $i$ we assume that 
$c_i=0$ ($c_i=1$) corresponds to the presence of a vacancy (particle) at the lattice-site $i$. The vector space of each lattice-site is 
two dimensional with the basis vector 
$\vert c_i=1\rangle = \left(\begin{array}{c}
    0 \\ 1
  \end{array}\right)$ 
and $\vert c_i=0\rangle =  \left(\begin{array}{c}
    1 \\ 0
  \end{array}\right)$. The configuration space of the system ${(\mathbbm{C}^2)}^{\otimes L}$ is $2^L$ dimensional. 
The modified Hamiltonian $\hat{\cal{H}}(s)$ for this current should be written as 
\begin{equation}
\label{hs}
\hat{\cal{H}}(s) =  \sum_{k=1}^{L-1}  {\cal I}^{\otimes (k-1)}\otimes  \hat{h}(s)  \otimes {\cal I}^{\otimes (L-k-1)}
\end{equation}
in which $\cal I$ is a $2\times 2 $ identity matrix in the basis $(0,1)$ and that
$$
\hat{h}(s) =\left(
\begin{array}{cccc}
 0 & 0 & 0 & 0 \\
 0 & -q (\Delta +1) & q^{-1}e^s & q^{-1}e^s \\
 0 & q e^{-s} & -q^{-1}(\Delta +1) & qe^{-s} \\
 0 & q \Delta e^{-s} & q^{-1} \Delta e^{s} & -q-q^{-1} \\
\end{array}
\right) 
$$
which is written in the basis $(00,01,10,11)$. 

In FIG.~\ref{fig1} we have plotted the numerically obtained largest eigenvalue of~(\ref{hs}) and its derivative and also the large deviation function 
of the current for a system of size $L=8$. The quantity $-d\Lambda^{\ast}(s)/ds$ at $s=0$ gives the average current in the steady-state $J^{\ast}$
which can be calculated using a matrix product method quite straightforwardly~\cite{HSP96,BE07}. At $s\neq 0$ the quantity $-d\Lambda^{\ast}(s)/ds$ gives  
the average of the current. At $s=s^{\ast}=\ln q^2$ the average current is $-J^{\ast}$.

Let us briefly review the basics of the matrix product method~\cite{BE07}. According to this method the steady-state probability distribution of a given 
configuration $c= \{ c_1, \cdots , c_L\}$ is given by
\begin{equation}
\label{NESS}
P^{\ast}(c)=  \langle c \vert P^{\ast} \rangle \propto \langle W \vert  \prod_{i=1}^{L}(c_i D+(1-c_i)E)   \vert V \rangle
\end{equation} 
in which $E$ and $D$ are non-commuting square matrices while $\langle W \vert$ and $ \vert V \rangle$ are vectors. These matrices and vectors
satisfy a quadratic algebra which has a four-dimensional matrix representation~\cite{HSP96}. Using~(\ref{NESS}) one can, in principle, calculate the average 
of any observable in the steady-state including the reaction-diffusion current explained before. This has actually been done in ~\cite{TJ13} hence the average 
current at $s^{\ast}$ is exactly known. 

In a separate work ~\cite{JM08} it has been found that the process defined by~(\ref{Rules}) has the following property: it has an invariant state-space 
under the evolution generated by $\hat{\cal{H}}(0)$ in~(\ref{hs}). This state-space consists of product shock measures with two shock fronts
at the lattice-sites $i$ and $j$ where $0\le i \le  j -1 \le L $. Note that the lattice-sites $0$ and $L+1$ are auxiliary lattice-sites to have a
well-defined shock measure. The structure of the product shock measure is 
\begin{equation}
  \label{measure}
  \vert i,j \rangle =
  \left(\begin{array}{c}
    1 \\ 0
  \end{array}\right)^{\otimes i} \otimes
  \left(\begin{array}{c}
    1-\rho \\ \rho
  \end{array}\right)^{\otimes j-i-1}
\otimes
  \left(\begin{array}{c}
    1 \\ 0
  \end{array}\right)^{\otimes L-j+1}
\end{equation}
in which $\rho=\Delta/(1+\Delta)$ so that 
\begin{equation}
\label{LS}
\hat{\cal{H}}(0)   \vert i,j \rangle = \sum_{i',j'}  \chi_{i',j'} \vert i',j' \rangle \; .
\end{equation}
The coefficients $\chi_{i',j'} $ are explicitly given in ~\cite{JM08}.  It has been shown that the shock fronts at two lattice-sites $i$ and $j$ perform 
simple random walk on the lattice. More precisely, while for $q>1$ the left shock front is always biased to the left, the bias of the 
right shock front depends on the values of both $q$ and $\Delta$. For $q^2 > 1+\Delta$ the right shock front is biased to the left
while for $q^2 < 1+\Delta$  the right shock front is biased to the right. Moreover, using~(\ref{LS}) the steady-state of 
the system can be  constructed as a linear superposition of the states of type~(\ref{measure}) with exactly known coefficients. 
It is clear that $\vert P^{\ast}\rangle=\vert \Lambda^{\ast}(s=0) \rangle$. 

It is not difficult to check that at $s^{\ast}$ we have
\begin{equation}
\label{Equs}
\begin{array}{lll}
\hat{\cal{H}}(s^{\ast}) \vert i,j \rangle = &q \vert i+1,j \rangle + q^{-1}(1+\Delta) \vert i-1,j \rangle& \\
&+q(1+\Delta)\vert i,j+1 \rangle+q^{-1} \vert i,j-1 \rangle & \\
&-(q+q^{-1})(2+\Delta) \vert i,j  \rangle & \\
& \quad \mbox{for}  \quad i=1,\cdots,L-2 & \\
& \quad \mbox{and}  \quad j=i+2,\cdots,L &\\  & \\
\hat{\cal{H}}(s^{\ast}) \vert 0,j \rangle= &q(1+\Delta)\vert 0,j+1 \rangle+q^{-1}  \vert 0,j-1 \rangle&\\
&-(q+q^{-1}(1+\Delta))\vert 0,j \rangle &\\
& \quad \mbox{for}  \quad j=2,\cdots,L &\\  & \\
\hat{\cal{H}}(s^{\ast}) \vert i,L+1 \rangle=& q\vert i+1,L+1 \rangle &\\
&+q^{-1}(1+\Delta)\vert i-1,L+1 \rangle &\\
&-(q^{-1}+q(1+\Delta))\vert i,L+1 \rangle &\\  
& \quad \mbox{for}  \quad i=1,\cdots,L-1 &\\  & \\
\hat{\cal{H}}(s^{\ast}) \vert i,i+1 \rangle& =0  \quad \mbox{for} \quad i=0,\cdots,L & \\
\hat{\cal{H}}(s^{\ast}) \vert 0,L+1 \rangle &=0 \; .& 
\end{array}
\end{equation}
These relations mean that as long as the shock fronts are far from the boundaries of the lattice i.e. $i\ne 0$ and $j\ne L+1$, they show 
the the same simple random walk behavior. However, as soon as the right  (left) shock front attaches to the right (left) 
boundary, it will not detach (reflect) from there. In other words the following product measure  
\begin{equation}
\vert 0,L+1 \rangle=  \left(\begin{array}{c}
    1-\rho \\ \rho
  \end{array}\right)^{\otimes L}
\end{equation}
is the right eigenvector of $\hat{\cal{H}}(s^{\ast})$ with zero eigenvalue that is we have $\vert \Lambda^{\ast}(s^{\ast}) \rangle=\vert 0,L+1 \rangle$.
We can also calculate the elements of the left eigenvector of $\hat{\cal{H}}(s^{\ast})$ as follows
\begin{equation}
 \langle \tilde{\Lambda}^{\ast}(s^{\ast})\vert c \rangle = \frac{\langle W \vert \prod_{i=1}^{L}(c_i D+(1-c_i)E)  \vert V \rangle}
 {Z \rho^{\sum_{i=1}^{L}c_i}(1-\rho)^{L-\sum_{i=1}^{L}c_i}}
\end{equation}
where the normalization factor $Z$ which is given by $\langle W \vert (D+E)^L \vert V \rangle$ can be calculated using the matrix representation
of the algebra given in~\cite{HSP96}.

From~(\ref{enantiodromy 1}) at $s=0$ one can easily see that $\hat{\cal{H}}_{\text{eff}}(s^{\ast})$ and $\hat{\cal{H}}_{\text{eff}}(0)$ have exactly the same 
right eigenvector with zero eigenvalue
$$
\vert P_{\text{ TTI}} (s=0) \rangle=\vert P_{\text{ TTI}} (s=s^{\ast}) \rangle \; .
$$

\begin{figure}
\includegraphics[scale=0.06]{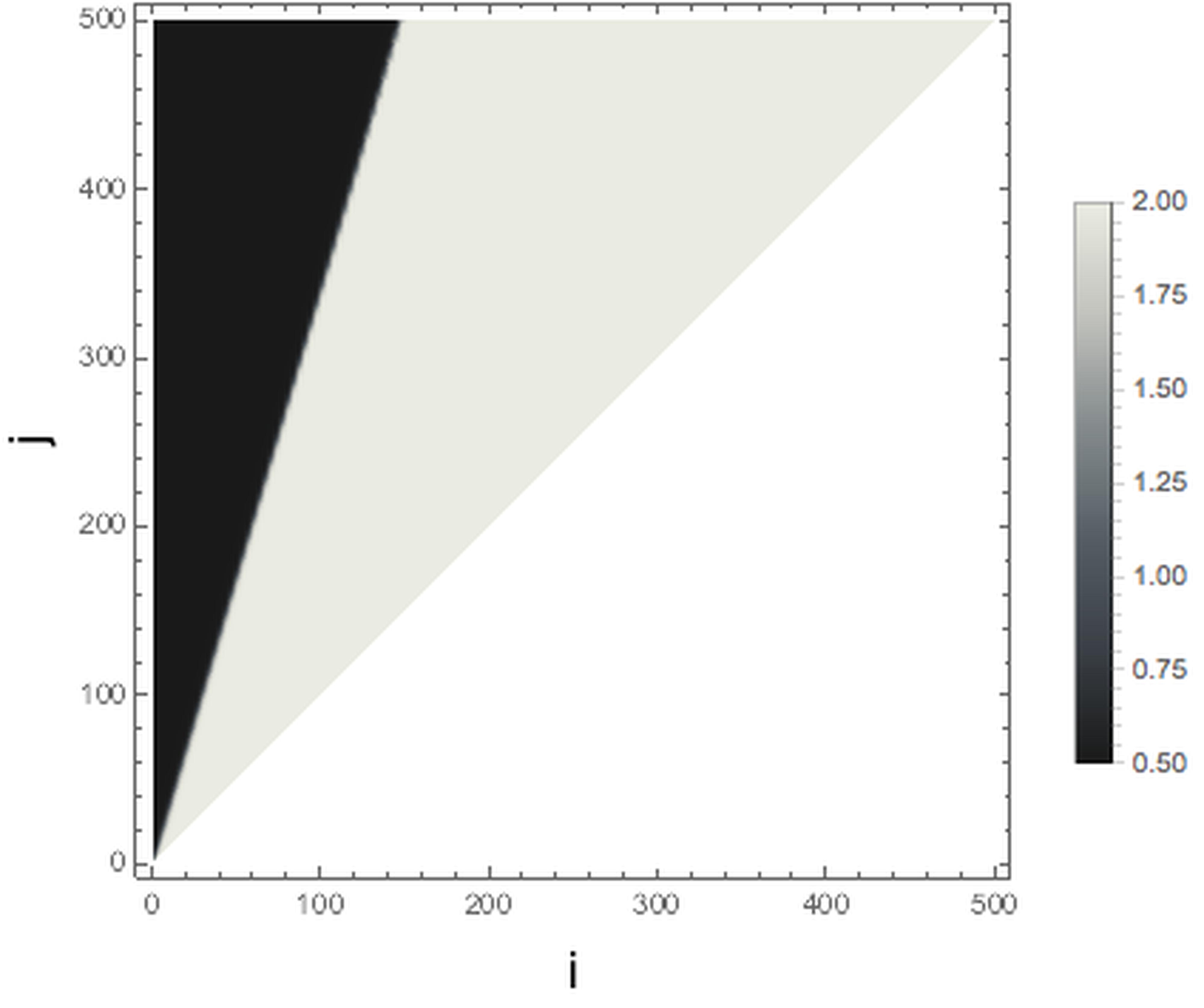}
\caption{\label{fig2} Density plot of~(\ref{final}) as a function of $i$ and $j$ for $L=500$, $q=2$ and $\Delta=0.5$.}
\end{figure}

Using~(\ref{s=0}) one can easily recognize that the effective transition rates at $s^\ast$ depend on both the initial and final configurations. 
On the other hand, they can be calculated exactly using the matrix product method explained earlier. For two arbitrary configurations 
$c=\{ c_1,\cdots,c_L\}$ and $c'=\{ c'_1,\cdots,c'_L\}$ we have 
\begin{equation}
\label{efferate}
\omega^{\text{eff}}_{c\to c'}(\ln q^2)=\omega_{c'\to c}\frac{\langle W \vert \prod_{i=1}^{L}(c'_i D+(1-c'_i)E)  \vert V \rangle}
{\langle W \vert \prod_{i=1}^{L}(c_i D+(1-c_i)E)  \vert V \rangle} \; .
\end{equation}
Because of the local interaction nature of the original process, $c$ and $c'$ are different from each other only in the configurations
of two consecutive lattice-sites (say $i$ and $i+1$) 
$$
\cdots c_ic_{i+1}\cdots \to \cdots c'_ic'_{i+1}\cdots \; .
$$
As we mentioned, $E$ and $D$ do not commute with each other. At the same time, neither the numerator nor denominator of the  
fraction appeared in~(\ref{efferate}) can be decomposed into productive factors so that the final result depend only on the local configurations. 
This means that the interactions in the effective process at $s^\ast$ are non-local yet the effective transition rates can be calculated exactly. 
In what follows we will give an explicit example by considering the initial and final configurations as 
$$
\textrm{\quad $c$ \quad}  \underbrace{\emptyset \cdots \emptyset}_{i-1}  A  \underbrace{\emptyset \cdots \emptyset}_{j-i-1}A
 \underbrace{\emptyset \cdots \emptyset}_{L-j}
$$
and 
$$
\textrm{\quad $c'$ \quad} \underbrace{\emptyset \cdots \emptyset}_{i}A \underbrace{\emptyset \cdots \emptyset}_{j-i-2}A 
\underbrace{\emptyset \cdots \emptyset}_{L-j}
$$
respectively. This indicates the diffusion of the leftmost particle to the right. One can now write
\begin{equation}
\label{final}
\omega^{\text{eff}}_{c\to c'}=q\frac{\langle W \vert E^{i}DE^{j-i-2}DE^{L-j}  \vert V \rangle}{\langle W \vert E^{i-1}DE^{j-i-1}DE^{L-j}   \vert V \rangle}\; .
\end{equation}
This expression can be calculated exactly using the matrix representation of the operators and vectors. It turns out that the final result 
depends on $i$ (or $i$ and $j$) explicitly. Since the mathematical expression is rather complicated, we have 
plotted~(\ref{final}) in FIG.~\ref{fig2} for the system in the low-density phase. 
%%%%%%%%%%%%%%%%%%%%%%%
%%%%%%%%%%%%%%%%%%%%%%
For $q=2$ the diffusion rate to the right in the original process is $0.5$. However, as it can be seen in FIG.~\ref{fig2} the diffusion rate 
to the right can be as large as $2$ depending on the lattice-site numbers $i$ and $j$. In contrast in the high-density phase where 
$q^2 < 1+\Delta$, the effective diffusion rate to the right is almost constant and independent of $i$ and $j$
except where $i$ and $j$ are very close to each other.

%%%%%%%%%%%%%%%%%%%%%%%%%%%%%%%%%%%%%%%%%%%%%%%%%%%%%%%%%%%%%%%%%%%%%%%%
\section{Concluding remarks}
%%%%%%%%%%%%%%%%%%%%%%%%%%%%%%%%%%%%%%%%%%%%%%%%%%%%%%%%%%%%%%%%%%%%%%%%
In this paper we have investigated the connection between the stochastic generators of effective processes corresponding to two specific atypical 
values of an entropic current with the Gallavotti-Cohen symmetry. For a specific family of processes we have shown that the effective stochastic 
generators at the points $s$ and $E-s$ are enantiodromic with respect to each other. 
These two generators have the same spectrum and the same steady-states; however, one of them generates the process 
backward in time with respect to the other one. It is important to note that the characteristics of the interactions corresponding to these points 
could be completely different as we have shown it in an exactly solvable example with non-conserving dynamics including coagulation and decoagulation 
processes on a one-dimensional lattice with reflecting boundaries. Although the original model includes nearest-neighbor interactions, we 
have shown that the adjoint process consists of non-local interactions. 

The Gallavotti-Cohen symmetry has been observed in the systems for which the characteristic polynomial 
of the modified Hamiltonian is symmetric with respect to some external field. However, there are also situations in which only 
the dominant eigenvalue of the modified Hamiltonian is symmetric. On the other hand, it has been shown that under some 
conditions on the structure of the configuration space and the reaction rates, the large deviation 
functions for the probability distributions of time-integrated currents satisfy a so called the Gallavotti-Cohen-like symmetry~\cite{BCHM12}. 
It would be interesting to investigate the connections between the effective stochastic generators corresponding to specific fluctuations 
in these cases.
%%%%%%%%%%%%%%%%%%%%%%%%%%%%%%%%%%%%%%%%%%%%%%%%%%%%%%%%%%%%%%%%%%%%%%%%
\section*{Acknowledgment}
We would like to thank the anonymous referee for his/her enlightening comments and criticisms that were very helpful to improve this manuscript.

%%%%%%%%%%%%%%%%%%%%%%%%%%%%%%%%%%%%%%%%%%%%%%%%%%%%%%%%%%%%%%%%%%%%%%%%

\end{document}